\documentclass[12pt]{article}
\usepackage{amsthm,amsmath,amsfonts,amssymb}
\usepackage{epsfig}
\usepackage{color}
\usepackage{graphicx}
\newcommand{\chib}{\overline{\chi}}

\newcommand{\HH}{\mathcal{H}}

\renewcommand{\ker}{{\rm Ker}}
\newcommand{\ran}{{\rm Ran}}

\newcommand{\ben}{\begin{displaymath}}
\newcommand{\een}{\end{displaymath}}
\newcommand{\beqn}{\begin{equation}}
\newcommand{\eeqn}{\end{equation}}
\newcommand{\beqna}{\begin{eqnarray*}}
\newcommand{\eeqna}{\end{eqnarray*}}

\def\ran{{\rm Ran} \, }

\newtheorem{lemma}{Lemma}
\newtheorem{theorem}[lemma]{Theorem}

\begin{document}
\author{M.~Griesemer and D.~Hasler}

\title{On the Smooth Feshbach-Schur Map}
\author{\vspace{5pt} M. Griesemer$^1$\footnote{
E-mail: marcel@mathematik.uni-stuttgart.de.} and D.
Hasler$^2$\footnote{E-mail: dh8ud@virginia.edu.} \\
\vspace{-4pt} \small{$1.$ Fachbereich Mathematik,
Universit\"at Stuttgart,} \\ \small{D-70569 Stuttgart, Deutschland}\\
\vspace{-4pt}
\small{$2.$ Department of Mathematics, University of Virginia,} \\
\small{Charlottesville, VA 22904-4137, USA}\\}
\date{}
\maketitle


\begin{abstract}
A new variant of the Feshbach map, called smooth Feshbach map, has
been introduced recently by Bach et al., in connection with the renormalization
analysis of non-relativistic quantum electrodynamics. We analyze
and clarify its algebraic and analytic properties, and we generalize
it to non-selfadjoint partition operators $\chi$ and $\chib$.
\end{abstract}

\section{Introduction}

For the spectral analysis of non-relativistic QED a
\emph{renormalization transform} was introduced in \cite{BFS1998a,
BFS1998b} that reduces an eigenvalue problem for the Hamiltonian
$H$ to an equal one for an effective Hamiltonian on a smaller
Hilbert space, or, more precisely, a Hamiltonian with fewer
degrees of freedom. The heart of this renormalization transform is
Schur's block-diagonalization of the Hamiltonian $H$ with respect
to the decomposition $\HH=P\HH\oplus \bar{P}\HH$ of the Hilbert
space $\HH$ induced by suitably chosen projections $P$ and
$\bar{P}=1-P$: assuming that $\bar{P}H\bar{P}$ is invertible on
$\bar{P}\HH$, the Hamiltonian $H$ is invertible if and only if its
Schur complement, or Feshbach map,
\begin{equation}\label{eq:FS}
  F_{P}(H) = PHP - PH\bar{P}\big(\bar{P}H\bar{P}\big)^{-1}\bar{P} HP,
\end{equation}
is invertible on $P\HH$ \cite{BFS1998a, BFS1998b}. Moreover, the
kernels of $H$ and $F_{P}(H)$ have equal dimensions. In the
renormalization analysis of Bach et al. the projection operator
$P$ is the spectral projection $\chi_{[0,\rho]}(H_f)$ of a
self-adjoint operator, $H_f$, the field energy.

In the recent beautiful paper \cite{bacchefrosig:smo} a novel,
\emph{smooth Feshbach map} $H=T+W\mapsto F_{\chi}(H,T)$ with
surprisingly nice algebraic properties is introduced. In
the definition of $F_{\chi}(H,T)$, commuting self-adjoint operators $\chi$ and
$\chib$ with $\chi^2+\chib^2=1$ and $[\chi,T]=0=[\chib,T]$ play the roles of $P$ and
$\bar{P}$. This allows one, in the application to QED, to choose
$\chi$ and $\chib$ as smooth functions of $H_f$, which avoids
technical problems that were caused by the non-differentiability
of the function $\chi_{[0,\rho]}$ defining $P$ in the
renormalization map based on \eqref{eq:FS}. Since $\chi$ and
$\chib$ need not be projections, there is no obvious interpretation of
the smooth Feshbach map in terms of a block-diagonalization of
$H$. Nevertheless, $H$ is invertible if and only if
$F_{\chi}(H,T)$ is invertible, the kernels of $H$ and
$F_{\chi}(H,T)$ have equal dimensions, and all other properties of
$F_P(H)$ that were used in \cite{BFS1998a, BFS1998b} have analogs
in the smooth Feshbach map. This is the content of the Feshbach
theorem, Theorem~II.1, in \cite{bacchefrosig:smo}.

In the present paper we prove that the Feshbach theorem is still
true when the self-adjointness assumption on $\chi$ and $\chib$ is
dropped. This generalization is needed, for example, in the analysis of
resonances, \cite{Fau2006}, and in our own forthcoming analysis of
the ground state. In the course of modifying the proof of Theorem
II.1, \cite{bacchefrosig:smo}, we closely examined all of its
parts. The result is a improved version of the Feshbach theorem,
Theorem~\ref{sfm} below,  with weaker assumptions and a stronger
statement. Using new algebraic identities, we show, for example,
that $\chi$ is an isomorphism from the kernel of $H$ onto the
kernel of $F_{\chi}(H,T)$, and we identify its inverse.

The Schur complement (for matrices) goes back to Schur
\cite{Schur1917}, see also \cite{Schur-memory, Zhang2005}, and it is widely
used in applied mathematics \cite{Zhang2005}. In the physics
literature H.~Feshbach derived an effective Hamiltonian of the
form of a Schur complement in a study of nuclear reactions
\cite{Feshbach1958}. Subsequently this effective Hamiltonian was
written in the form \eqref{eq:FS} using projection operators $P$
and $Q=1-P$ \cite{hahmalspr1962}, called Feshbach's projection operators \cite{mal1966}.

\emph{Acknowledgments.} M. Griesemer thanks Ira Herbst for the hospitality at
the University of Virginia, where most of this work was done and Arne Jensen
for pointing out the references to the original work of Schur. We thank Joseph
H.~Macek for the references \cite{hahmalspr1962, mal1966} .

\section{The Smooth Feshbach Map}

Let $\chi$ and $\overline{\chi}$ be commuting, nonzero bounded
operators, acting on a separable Hilbert space $\HH$ and
satisfying $\chi^2 + \overline{\chi}^2 = 1$. By a \textbf{Feshbach
pair} $(H,T)$ for $\chi$ we mean a pair of closed operators with
same domain
$$
    H,T: D(H)=D(T)\subset \HH\to\HH
$$
such that $H,T,W:=H-T$, and the operators
\begin{align*}
W_\chi &:= \chi W \chi,& W_{\chib} &:= \chib W \chib, \\
H_\chi &:= T + W_\chi,& H_{\chib} &:= T +  W_{\chib},
\end{align*}
defined on $D(T)$ satisfy the following assumptions:
\begin{itemize}
\item[(a)] $\chi T\subset T\chi$ and $\chib T\subset T\chib$,
\item[(b)] $T, H_{\chib}:D(T)\cap\ran\chib\to \ran\chib$ are bijections with bounded inverse.
\item[(c)] $\chib H_{\chib}^{-1} \chib W \chi: D(T)\subset\HH\to\HH$ is a bounded operator.
\end{itemize}

Henceforth we will call an operator $A: D(A) \subset \HH \to \HH$
\emph{bounded invertible in a subspace} $V \subset \HH$ ($V$ not
necessarily closed), if $A : D(A) \cap V \to V$ is a bijection
with bounded inverse.

\noindent\textbf{Remarks.}
\begin{enumerate}
\item To verify (a), it suffices to show that $T\chi=\chi T$ and $T\chib=\chib
T$ on a \emph{core} of $T$.

\item If $T$ is bounded invertible in $\ran\chib$, $\|T^{-1}\chib
W\chib\|<1$ and $\|\chib WT^{-1}\chib\|<1$, then the bounded
invertibility of $H_{\chib}$ and condition (c) follow. See
Lemma~\ref{lm:F-cond} below.

\item Note that $\ran\chi$ and $\ran\chib$ need not be closed and
are not closed in the application to QED. One can however, replace
$\ran\chib$ by $\overline{\ran\chib}$ both in condition (b) and in
the statement of Theorem~\ref{sfm}, below. Then this theorem continues to
hold and the proof remains unchanged.
\end{enumerate}

Since our conditions defining Feshbach pairs are different from those stated in 
\cite{bacchefrosig:smo}, some explanations are necessary.
First, our conditions (a) and (b) on Feshbach pairs can also be found in  
\cite{bacchefrosig:smo}, Section~2.1. The bounded invertibility of $T$ is not
mentioned there as an assumption, but it is used in
the proof of Theorem~2.1, \cite{bacchefrosig:smo}. Second, there is no condition needed on $\chi
W(\chib H_{\chib}^{-1} \chib)$, or a similar operator, since this operator 
is bounded as a consequence of the domain assumptions.  
In fact, since $H$ and $T$ are closed on $D(T)$, and since $\ran \chib
H_{\chib}^{-1}\chib\subset D(T)$, the operators $H(\chib
H_{\chib}^{-1} \chib)$, $T(\chib H_{\chib}^{-1}\chib)$ are defined on $\HH$, closed
and hence bounded. Since $W=H-T$, it follows that $W(\chib
H_{\chib}^{-1}\chib)$ is bounded. Third, our condition (c) is weaker than the corresponding condition (2.3) of 
\cite{bacchefrosig:smo}, at least in practice, and this is crucial in
some applications to QED. Condition (c) is satisfied, for example, if $H=H_{\alpha}$ is
the Hamiltonian of an atom or molecule in the standard model of
non-relativistic QED with finestructure constant $\alpha$ and with
$T=H_{\alpha=0}$. Condition (2.3) of \cite{bacchefrosig:smo} will not be satisfied in this case. 

Given a Feshbach pair $(H,T)$ for $\chi$, the operator
\begin{equation}\label{eq:fesh}
    F_\chi(H,T) := H_{\chi} - \chi W \chib H_{\chib}^{-1} \chib W \chi
\end{equation}
on $D(T)$ is called \emph{Feshbach map of} $H$. The mapping
$(H,T)\mapsto F_\chi(H,T)$ is called \emph{Feshbach map}. The
auxiliary operators
\begin{eqnarray*}
\begin{array}{ll}
Q_\chi := \chi - \chib  H_{\chib}^{-1} \chib W \chi  \\
Q_\chi^\# := \chi - \chi W \chib H_{\chib}^{-1} \chib,
\end{array}
\end{eqnarray*}
play an important role in the analysis of $F_\chi(H,T)$. By
conditions (a), (c), and the explanation above, they are bounded, and
$Q_{\chi}$ leaves $D(T)$ invariant. The Feshbach map is
isospectral in the sense of the following Theorem. It generalizes
Theorem 2.1 in \cite{bacchefrosig:smo} non-selfadjoint $\chi$ and
$\chib$.

\begin{theorem}\label{sfm} Let $(H,T)$ be
a Feshbach pair for $\chi$ on a separable Hilbert space $\HH$.
Then the following holds:
\begin{enumerate}
\item[(i)] Let $V$ be a subspace with  $\ran \chi \subset V
\subset \HH$,
\begin{equation} \label{eq:feshsubspace} T: D(T) \cap V \to V,\qquad{\rm  and}\qquad\chib T^{-1} \chib V\subset V \; .
\end{equation}
Then $H:D(H)\subset\HH \to \HH$ is bounded invertible if and only
if $F_\chi(H,T): D(T) \cap V \to V$ is bounded invertible in $V$.
Moreover,
\begin{eqnarray*}
   H^{-1} &=& Q_{\chi}F_{\chi}(H,T)^{-1} Q_{\chi}^{\#} + \chib H_{\chib}^{-1}\chib,\\
    F_\chi(H,T)^{-1} &=& \chi H^{-1}\chi + \chib T^{-1}\chib.
\end{eqnarray*}

\item[(ii)] $\chi \ker H
\subset \ker F_{\chi}(H,T)$ and $Q_{\chi}\ker F_{\chi}(H,T)\subset\ker H$.
The mappings
\begin{align}
\chi :&  \  \ker H \to \ker F_\chi(H,T)  \label{eq:keriso1}, \\
Q_\chi :&  \ \ker F_\chi(H,T) \to \ker H  \label{eq:keriso2},
\end{align}
are linear isomorphisms and inverse to each other.
\end{enumerate}
\end{theorem}

\noindent\textbf{Remarks.}
\begin{enumerate}
\item The subspaces $V = \ran \chi$ and $V = \HH$ satisfy the
conditions stated in \eqref{eq:feshsubspace}.
\item From \cite{bacchefrosig:smo} it is known that $\chi$ and $Q_{\chi}$
are one-to-one on $\ker H$ and $\ker F_{\chi}(H,T)$ respectively. The
stronger result (ii) will be derived from the new algebraic
identities (a) and (b) of the following lemma.
\end{enumerate}

Theorem~\ref{sfm} will easily follow from the next lemma, which is of
interest and importance in its own right.

\begin{lemma}\label{lm:F-basics}
Let $(H,T)$ be a Feshbach pair for $\chi$ and let
$F:=F_{\chi}(H,T)$, $Q := Q_\chi$, and $Q^{\#} := Q^{\#}_\chi$ for
simplicity. Then the following identities hold:
\begin{align*}
(a)&& (\chib H_{\chib}^{-1} \chib)H &= 1-Q\chi,\quad\text{on}\
D(T),&\qquad
H(\chib H_{\chib}^{-1} \chib) &= 1-\chi Q^{\#},\quad \text{on}\ \HH,\\
(b)&& (\chib T^{-1}\chib)F &= 1-\chi Q,\quad\text{on}\
D(T),&\qquad
F(\chib T^{-1}\chib)&=1-Q^{\#}\chi,\quad \text{on}\ \HH,\\
(c)&& HQ &=\chi F,\quad\text{on}\ D(T),&\qquad Q^{\#}H &= F\chi,
\quad\text{on}\ D(T).\\
\end{align*}
\end{lemma}

\begin{proof}
We proof the first equations in (a), (b), and (c) only. The other
ones are proved analogously. (a) Since $\chib T\subset \chib T$
and $\chi^2+\chib^2=1$, on $D(T)$,
\begin{eqnarray*}
  (\chib H_{\chib}^{-1} \chib)H  &=& \chib H_{\chib}^{-1} T\chib +
  \chib H_{\chib}^{-1} \chib W(\chi^2+\chib^2)\\
  &=& \chib H_{\chib}^{-1} (T+W_{\chib})\chib + \chib
  H_{\chib}^{-1}\chib W\chi^2\\
  &=& \chib^2 + \chib H_{\chib}^{-1}\chib W\chi^2\\
  &=& 1-Q\chi.
\end{eqnarray*}
(b) Using  again condition (a) of Feshbach pairs and
$\chi^2+\chib^2=1$, we find on $D(T)$,
\begin{eqnarray*}
(\chib T^{-1}\chib) F &=& \chib T^{-1} \chib ( T +  W_\chi - \chi
W \chib
H_{\chib}^{-1} \chib W \chi) \\
&=& {\chib}^2 + \chib T^{-1} \chib \chi W \chi - \chi \chib T^{-1}
W_{\chib} H_{\chib}^{-1} \chib W \chi \\
&=& {\chib}^2 + \chi \chib H_{\chib}^{-1} \chib W \chi \\
&=& 1 - \chi Q \; ,
\end{eqnarray*}
where in the third equation we used the resolvent identity $\chib
( T^{-1} - H_{\chib}^{-1} ) \chib = \chib T^{-1} W_{\chib}
H_{\chib}^{-1} \chib$.


\noindent
 (c) By the second equation of (a), on $D(T)$,
\begin{eqnarray*}
  HQ &=& H(\chi - \chib  H_{\chib}^{-1} \chib W \chi)\\
            &=& \chi T + W\chi - (1-\chi Q^{\#})W\chi\\
            &=& \chi(T+Q^{\#}W\chi)\ =\ \chi F
\end{eqnarray*}
\end{proof}

\noindent\textbf{Remark.} Alternatively, one can prove the
identities of Lemma \ref{lm:F-basics} (b) as follows. By
definition of $F$ and the first equation of (c), on $D(T)$,
\begin{eqnarray*}
\chib^2 F = F - \chi^2 F &=& (T + \chi W Q) - \chi H Q\\
   &=& T - \chi T Q\ =\ T(1-\chi Q).
\end{eqnarray*}
Since the range of $1-\chi Q = \chib^2 + \chi\chib
H_{\chib}^{-1}\chib W\chi$ is a subspace of $\ran\chib$, the first
identity of Lemma \ref{lm:F-basics} (b) follows from condition (b)
of Feshbach pairs. The other identity of Lemma \ref{lm:F-basics}
(b) can be shown similarly.

\begin{proof}[Proof of Theorem~\ref{sfm}]
We use the simplified notation of Lemma~\ref{lm:F-basics}.

(i) Suppose $F$ is bounded invertible in $V$. Then the operator
$$R :=  Q F^{-1} Q^\# +
\overline{\chi} H_{\overline{\chi}}^{-1} \overline{\chi}$$ is
bounded, and by Lemma~\ref{lm:F-basics} (a) and (c)
\begin{eqnarray*}
    RH &=& Q F^{-1}Q^{\#}H + (\chib H_{\chib}^{-1} \chib)H\\
       &=& Q\chi + (1-Q\chi)\ =\ 1,
\end{eqnarray*}
on $D(H)$. Similarly one shows that $HR=1$ on $\HH$. On the other
hand, if $H$ is bounded invertible in $\HH$, then
$$
\widetilde{R} := \chi H^{-1} \chi + \overline{\chi}
T^{-1}\overline{\chi} \;
$$
is bounded, and by Lemma~\ref{lm:F-basics} (c) and (b)
\begin{eqnarray*}
    \widetilde{R}F &=& \chi H^{-1}\chi F + (\chib
    T^{-1}\chib)F\\
    &=& \chi Q + (1-\chi Q)\ =\ 1,
\end{eqnarray*}
on $D(T)$. Similarly one shows that $F \widetilde{R} =1$ on $\HH$.
This shows that $F$ is bounded invertible in $\HH$. Finally, from
the definitions of $F$, $\widetilde{R}$ and the properties of $V$,
it follows that $F : D(T) \cap V \to V$ and $\widetilde{R}: V \to
D(T) \cap V$. Hence $F$ is also bounded invertible in $V$.

(ii) On the one hand, by Lemma~\ref{lm:F-basics} (c), $\chi \ker H
\subset \ker F$ and $Q \ker F \subset \ker H$. On the other hand,
by the first equations of part (a) and (b) of that lemma
$$
Q \chi = 1 \quad {\rm on } \ \ \ker H  \qquad {\rm and } \qquad
\chi Q = 1 \quad {\rm on } \ \ \ker F \; .
$$
This proves statement (ii).
\end{proof}

\begin{lemma}\label{lm:F-cond}
Conditions (a),(b), and (c) on Feshbach pairs are satisfied if
\begin{itemize}
\item[(a')] $\chi T\subset T\chi$ and $\chib T\subset T\chib$,
\item[(b')] $T$ is bounded invertible in $\ran\chib$,
\item[(c')] $\|T^{-1}\chib W\chib\|<1$ and $\|\chib WT^{-1}\chib\|<1$.
\end{itemize}
\end{lemma}

\begin{proof}
By assumptions (a') and (b'), on $D(T)\cap\ran\chib$
$$
    H_{\chib} = (1+\chib WT^{-1}\chib)T
$$
and $T:D(T)\cap\ran\chib\to\ran\chib$ is a bijection with bounded
inverse. From (c') it follows that
$$
   1+\chib WT^{-1}\chib : \ran\chib \to\ran\chib
$$
is a bijection with bounded inverse. In fact, $(1+\chib W
T^{-1}\chib)\ran\chib\subset\ran\chib$, the Neumann series
$$
     \sum_{n\geq 0} (-\chib WT^{-1}\chib)^n = 1-\chib W T^{-1}\chib\sum_{n\geq 0}
     (-\chib WT^{-1}\chib)^n
$$
converges and maps $\ran\chib$ to $\ran\chib$. Hence
$H_{\chib}\upharpoonright\ran\chib$ is bounded invertible.

Finally, from $H_{\chib}=T(1+T^{-1}\chib W{\chib})$ and (c') it
follows that
$$
    H_{\chib}^{-1}\chib W = (1+T^{-1}W_{\chib})^{-1}T^{-1}\chib W,
$$
which, by (c'), is bounded bounded.
\end{proof}


\begin{thebibliography}{10}

\bibitem{bacchefrosig:smo}
Volker Bach, Thomas Chen, J{\"u}rg Fr{\"o}hlich, and Israel~Michael Sigal.
\newblock Smooth {F}eshbach map and operator-theoretic renormalization group
  methods.
\newblock {\em J. Funct. Anal.}, 203(1):44--92, 2003.

\bibitem{BFS1998a}
Volker Bach, J{\"u}rg Fr{\"o}hlich, and Israel~Michael Sigal.
\newblock Quantum electrodynamics of confined nonrelativistic particles.
\newblock {\em Adv. Math.}, 137(2):299--395, 1998.

\bibitem{BFS1998b}
Volker Bach, J{\"u}rg Fr{\"o}hlich, and Israel~Michael Sigal.
\newblock Renormalization group analysis of spectral problems in quantum field
  theory.
\newblock {\em Adv. Math.}, 137(2):205--298, 1998.

\bibitem{Fau2006}
Jeremy Faupin.
\newblock Resonances of the confined hydrogenoid ion and the {D}icke effect in
  non-relativistic quantum electrodynamics.
\newblock mp-arc, 06-344.

\bibitem{Feshbach1958}
Herman Feshbach.
\newblock Unified theory of nuclear reactions.
\newblock {\em Ann. Physics}, 5:357--390, 1958.

\bibitem{hahmalspr1962}
Yukap Hahn, Thomas~F. O'Malley, and Larry Spruch.
\newblock Static approximation and bounds on single-channel phase shifts.
\newblock {\em Phys. Rev. (2)}, 128:932--942, 1962.

\bibitem{Schur-memory}
Anthony Joseph, Anna Melnikov, and Rudolf Rentschler, editors.
\newblock {\em Studies in memory of {I}ssai {S}chur}, volume 210 of {\em
  Progress in Mathematics}.
\newblock Birkh\"auser Boston Inc., Boston, MA, 2003.
\newblock Papers from the Paris Midterm Workshop of the European Community
  Training and Mobility of Researchers (TMR) Network held in Chevaleret, May
  21--25, 2000 and the Schur Memoriam Workshop held in Rehovot, December
  27--31, 2000.

\bibitem{mal1966}
T.~F. O'Malley.
\newblock Theory of dissociative attachment.
\newblock {\em Phys. Rev.}, 150(1):14--29, Oct 1966.

\bibitem{Schur1917}
J.~Schur.
\newblock {\"U}ber {P}otenzreihen die im {I}nneren des {E}inheitskreises
  beschr\"ankt sind.
\newblock {\em J. reine u. angewandte Mathematik}, pages 205--232, 1917.

\bibitem{Zhang2005}
Fuzhen Zhang, editor.
\newblock {\em The {S}chur complement and its applications}, volume~4 of {\em
  Numerical Methods and Algorithms}.
\newblock Springer-Verlag, New York, 2005.

\end{thebibliography}

\end{document}